# Current-induced magnetization reversal in (Ga,Mn)(Bi,As) epitaxial layer with perpendicular magnetic anisotropy

Tomasz Andrearczyk [1], Janusz Sadowski [1,2], Krzysztof Dybko [1,3], Tadeusz Figielski, [1] and Tadeusz Wosinski [1]*

[1] *Institute of Physics, Polish Academy of Sciences, PL-02668 Warsaw, Poland*
[2] *Department of Physics and Electrical Engineering, Linnaeus University, SE-391 82 Kalmar, Sweden*
[3] *International Research Centre MagTop, Institute of Physics, Polish Academy of Sciences, PL-02668 Warsaw, Poland*

**Abstract:**

Pulsed current-induced magnetization reversal is investigated in the layer of (Ga,Mn)(Bi,As) dilute ferromagnetic semiconductor (DFS) epitaxially grown under tensile misfit strain causing perpendicular magnetic anisotropy in the layer. The magnetization reversal, recorded through measurements of the anomalous Hall effect, appearing under assistance of a static magnetic field parallel to the current, is interpreted in terms of the spin-orbit torque mechanism. Our results demonstrate that an addition of a small fraction of heavy Bi atoms, substituting As atoms in the prototype DFS (Ga,Mn)As and increasing the strength of spin-orbit coupling in the DFS valence band, significantly enhances the efficiency of current-induced magnetization reversal thus reducing considerably the threshold current density necessary for the reversal. Our findings are of technological importance for applications to spin-orbit torque-driven nonvolatile memory and logic elements.

______________________

* Corresponding author e-mail: wosin@ifpan.edu.pl

## 1. Introduction

Relativistic interaction between the electron's spin and its angular momentum, known as spin-orbit coupling (SOC), gives rise to a number of phenomena in magnetic materials, including anomalous and planar Hall effects and spin-orbit torque [1-3]. The latter one, appearing in crystals or structures with broken inversion symmetry, can be utilized for electrically induced switching magnetization direction in ferromagnetic domains, which is of special interest for applications to spintronic devices [4]. Experimental demonstration of the effect was first reported for epitaxial layers of the prototype dilute ferromagnetic semiconductor (DFS) (Ga,Mn)As with in-plane magnetic anisotropy owing to the compressive misfit strain [5]. Later on, it was confirmed by other authors for similar layers [6-9], as well as for the (Ga,Mn)As layers grown under tensile misfit strain resulting in perpendicular magnetic anisotropy [7,10], and extended to metallic structures such as heavy-metal/ferromagnet heterostructures, c.f. [11-13].

In crystals with the bulk inversion asymmetry, e.g. with zinc-blende structure, the spin degenerate electron states split, even in the absence of a magnetic field, and this spin splitting depends on the electron wavevector **k**, as predicted by the Dresselhaus effect [14]. Symmetry reduction owing to uniaxial strain in the crystal causes additional spin-orbit splitting of the Dresselhaus-type, linear in $k$. In the presence of electric current an uneven occupation of +**k** and −**k** states appears and the charge carriers become partly spin-polarized thus generating a torque or effective magnetic field $\mathbf{B}_{SO}$ (different from the current-induced Oersted field) acting on the magnetization. The sign of this field depends on the sign of strain and its magnitude increases with the strain magnitude [7]. Moreover, Bychkov and Rashba [15] have shown that layered crystal structures with a single high-symmetry axis normal to the layer plane are also subject to the spin splitting linear in $k$. Both the Dresselhaus and Rashba spin-orbit-induced effective magnetic fields interact with localized magnetic ions present in the crystals thus leading to the magnetization switching.

As the SOC strength is generally enhanced in materials composed of heavy elements we have grown and investigated the quaternary (Ga,Mn)(Bi,As) compound containing a small fraction of heavy Bi atoms, replacing As atoms in (Ga,Mn)As [16,17]. The epitaxial (Ga,Mn)(Bi,As) layers containing 1% Bi exhibit homogeneous ferromagnetic ordering below the Curie temperature, similar to that in the reference layers without the Bi content [18], and, as a result of enhanced SOC strength, significantly increased magnitudes of the anisotropic magnetoresistance [19] and planar Hall effect [20].

## 2. (Ga,Mn)(Bi,As) layer growth and characterization

In the present study we have investigated 15 nm thick (Ga,Mn)(Bi,As) DFS layer, with 6% Mn and 1% Bi contents, grown by the low-temperature molecular-beam epitaxy (MBE) technique at the temperature of approximately 230°C on semi-insulating (001)-GaAs substrate covered with a 0.63 μm thick, fully plastically relaxed, $In_{0.2}Ga_{0.8}As$ buffer layer. In-situ reflection high-energy electron diffraction (RHEED) has been used to verify the DFS layer thickness and Mn composition [19,21]. After the growth the wafer has been subjected to a low-temperature annealing treatment in air at 180°C for 50 h to improve the magnetic properties of the DFS layer through out-diffusion of Mn interstitials, cf. [16,22]. As a result of the use of InGaAs buffer the DFS layer exhibits an in-plane biaxial tensile misfit strain of about 1%, as determined from the high-resolution X-ray diffractometry results presented in our previous paper for the 50 nm thick layer grown under the same conditions [19].

Magneto-electric properties of the DFS layer have been investigated employing simple Hall-bars of about 1.5 mm width, cleaved along the [1$\bar{1}$0] crystallographic direction from the wafer. A picture of the sample, whose results are presented in this letter, supplied with indium-soldered Ohmic contacts to the DFS layer, is shown in Fig. 1. Four-probe longitudinal resistance $R_{xx}$ and anomalous Hall resistance $R_{xy}$ of the Hall-bars have been measured using a 1 μA sensing current and the Keithley Delta mode technique. A rotating helium cryostat has been used to apply an external magnetic field at either perpendicular or in-plane directions.

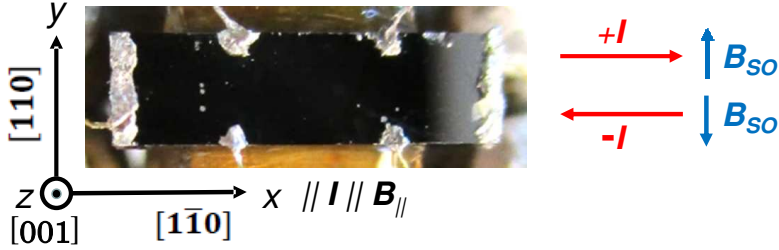

**Fig. 1**. Sample configuration for the magnetotransport measurements and the dependence of spin-orbit-induced effective magnetic field $\mathbf{B}_{SO}$ on the sign of current $I$ flowing along the Hall-bar of tensile-strained (001) layer of zinc-blende structure; in agreement with Refs. 7 and 10. $B_∥$ is an external in-plane magnetic field parallel to the current direction.

The ferromagnetic Curie temperature $T_C$ of the DFS layer can be evaluated from temperature dependence of the Hall-bar longitudinal resistance, shown in Fig. 2. As the position of broad maximum in that dependence suggests rather high value of $T_C$, much above 100 K, a maximum on the dependence of temperature derivative of longitudinal resistance on temperature, also shown in Fig. 2, determines more precisely the $T_C$ value at 130 K [23].

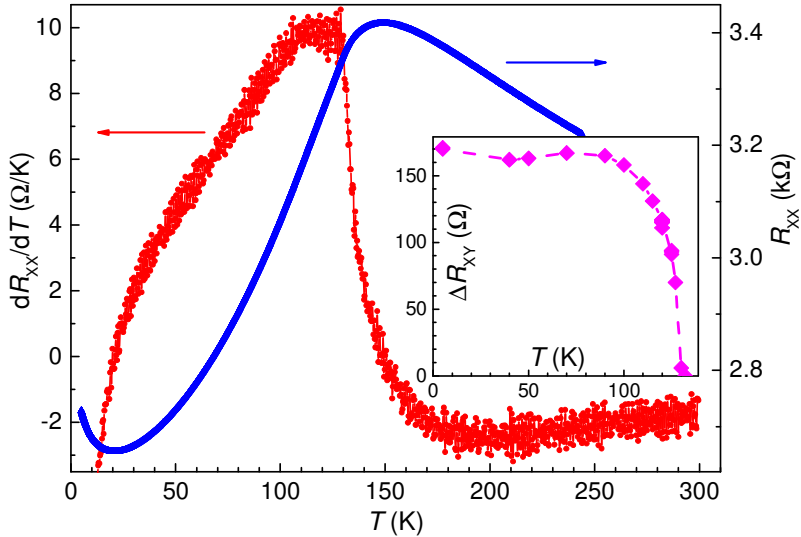

**Fig. 2**. Temperature dependences of longitudinal resistance and the derivative of that resistance with respect to temperature for the (Ga,Mn)(Bi,As) Hall-bar. Inset presents temperature dependence of the magnitude of anomalous Hall resistance hysteresis loop.

Anomalous Hall resistance of a ferromagnetic layer is proportional to the vertical component of its magnetization $M_z$. While sweeping up and down an external magnetic field applied perpendicular to the layer plane, $B_⊥$, the anomalous Hall resistance of the investigated

Hall-bar displays a rectangular hysteresis loop, similar to the magnetization hysteresis loop of the layer, as shown in Fig. 3 for various temperatures. It can be used to monitor the magnetization direction in the bar and its temperature dependence. These results prove the existence of perpendicular magnetic anisotropy in the DFS layer with the easy magnetization axis along the out-of-plane [001] crystallographic direction, in agreement with the results of superconducting quantum interference device (SQUID) magnetometry measurements performed for similarly grown (Ga,Mn)(Bi,As) layers [18]. The temperature dependence of magnitude of the $R_{xy}$ hysteresis loop, shown in the inset in Fig. 2, confirms $T_C \approx 130$ K.

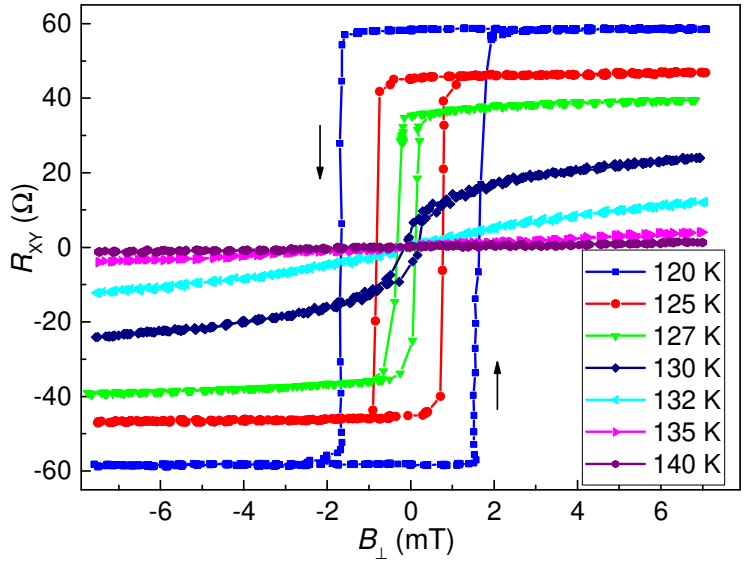

**Fig. 3**. Anomalous Hall resistance measured for the (Ga,Mn)(Bi,As) Hall-bar at various temperatures, written in the figure, while sweeping an external magnetic field, perpendicular to the layer plane, in opposite directions, as indicated by the arrows.

## 3. Current-induced magnetization reversal

Applying a single pulse of electric current along the Hall-bar, assisted by a static in-plane magnetic field along the bar length, we have been able to reverse the magnetization direction in the DFS layer. In Fig. 4 we demonstrate the magnetization switching recorded through measuring the anomalous Hall resistance of the Hall-bar, $R_{xy}$, at the temperature of 120 K. For the results shown in Fig. 4a, a positive magnetic field perpendicular to the layer plane was applied before the measurement to align the initial magnetization in the layer along the $+z$ direction. After decreasing this field to zero, an external in-plane magnetic field along the $+x$ direction, $B_{\parallel}$, has been applied. Injecting a negative current pulse of the intensity of 13 mA, corresponding to the current density of $5.8 \times 10^4$ A/cm$^2$, and duration of 36 ms, schematically shown in Fig. 4b, causes the reversal of magnetization by 180°. Another current pulse of opposite sign restores the initial magnetization direction in the layer. During applying the current pulses, whose intensity is over 4 orders of magnitude larger than that of measuring current, the configuration of experimental setup has to be changed and the $R_{xy}$ resistance is not registered (dashed lines in Figs. 4a and 4c). On the other hand, the sample temperature is monitored throughout the whole experiment showing negligible temperature fluctuations (below ±0.2 K) due to possible Joule heating during the current pulses.

At the temperature of 120 K the $R_{xy}$ resistance varies within approximately ±60 Ω, which is consistent with the magnitude of the $R_{xy}$ hysteresis loop, shown in Fig. 3, indicating that the magnetization is fully reversed between $+M_z$ and $-M_z$ directions by the current. Under the positive in-plane magnetic field (Fig. 4a) the pulse of negative current switches the magnetization to $-M_z$ and that of positive current results in the opposite switching. When the in-plane field is reversed the action of current pulses reverses (Fig. 4c). The value of applied in-plane magnetic field of 14 mT is about one order of magnitude smaller than the one, which causes the magnetization reversal in the absence of a current pulse, as shown with the red curve in Fig. 5. This reversal appears because of not ideal in-plane direction of that field, which contains a small perpendicular component. By comparing the field, at which the reversal occurs, with the value of coercive field corresponding to the magnetization reversal under perpendicular field (blue curve in Fig. 5), we can determine the nonideality angle to be about 1° towards the $-M_z$ direction in our experiments.

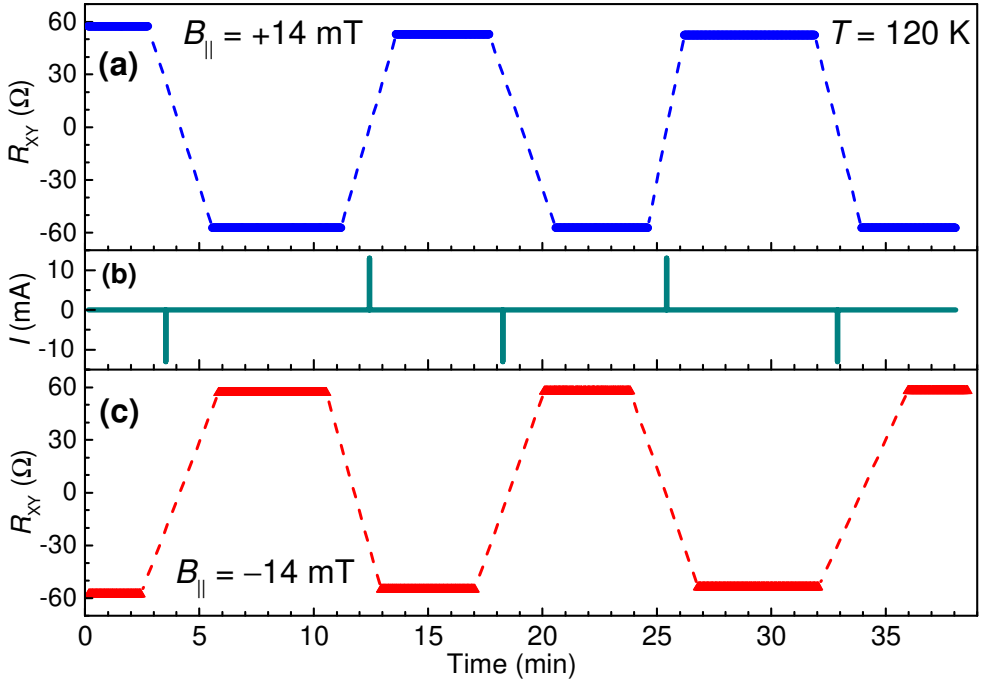

**Fig. 4**. Current-pulse-induced magnetization switching in the (Ga,Mn)(Bi,As) Hall-bar at the temperature 120 K measured by the anomalous Hall resistance $R_{xy}$ as a function of time. (a) $R_{xy}$ measured under the positive parallel magnetic field $B_{\parallel} = 14$ mT after magnetization of the Hall-bar under positive perpendicular magnetic field of 0.5 T. (b) Schematic of the sequence of current pulses of 13 mA and 36 ms duration. (c) Same as (a) but for the negative parallel magnetic field, after magnetization of the Hall-bar under negative perpendicular magnetic field. Dashed lines in (a) and (c) denote periods when $R_{xy}$ has not been measured.

The revealed bipolar reversal of perpendicular magnetization induced by injecting in-plane current pulse, occurring under assistance of a static magnetic field parallel to the current, corresponds to one of the switching schemes considered for the spin-orbit torque mechanism [11,13]. Our results can be interpreted in terms of the effective magnetic field perpendicular to both the parallel external field and current-induced spin-orbit field; c.f. [11]. This interpretation is consistent with the theoretical model proposed by Engel et al. [24] showing that the combination of parallel magnetic field and current-induced spin-orbit field generates an out-of-plane spin polarization of current-induced carriers proportional to the vector product of $\mathbf{B}_{\parallel} \times \mathbf{B}_{SO}$. The sign of this operation is reversed while changing the sign of either external magnetic field or spin-orbit field (by changing the sign of current pulse in our experiment), in agreement with our experimental results.

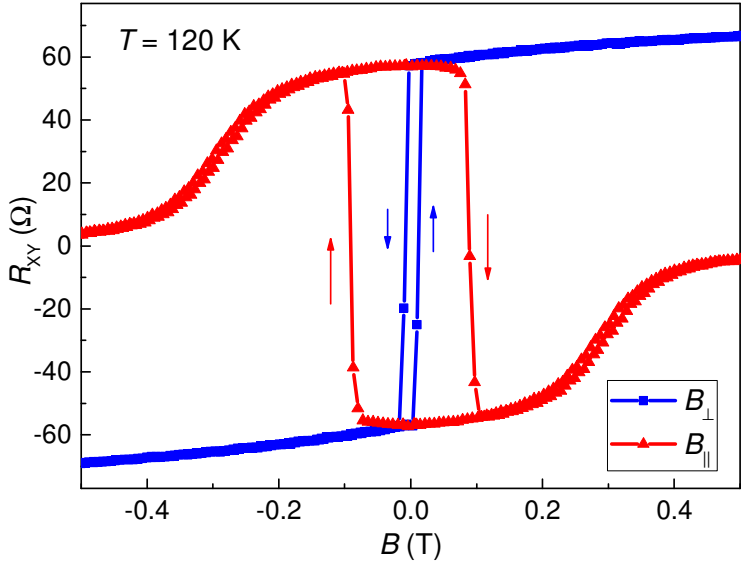

**Fig. 5**. Anomalous Hall resistance measured for the (Ga,Mn)(Bi,As) Hall-bar at the temperature 120 K while sweeping an external magnetic field, perpendicular to the Hall-bar plane (squares and blue curve) and parallel to the Hall-bar length (triangles and red curve), in opposite directions, as indicated by the arrows.

The current density threshold for switching magnetization in the (Ga,Mn)(Bi,As) Hall-bar is about $5\times10^4$ A/cm$^2$ in our experiments (at the temperature of 120 K and pulse width of 36 ms), which is almost three orders of magnitude smaller than that in metallic ferromagnets [11,13] and about 6 times smaller than the one necessary for magnetization reversal in tensile strained (Ga,Mn)As layer containing the same fraction on Mn ions [10]. Low switching current density is beneficial for applications to power efficient spintronic devices, especially important as the spin-orbit torque-driven magnetization manipulation is expected to be applied in next generation non-volatile magnetic random-access memories (MRAMs) and non-volatile logic elements [4,25]. In those applications the spin-orbit torque mechanism offers several advantages over the currently exploited spin-transfer torque one, such as reduced power consumption and faster device operation.

## 4. Conclusions

Bipolar reversal of the magnetization in (Ga,Mn)(Bi,As) layer, achieved by injecting in-plane current pulses, of either positive or negative sign, parallel to static magnetic field of moderate amplitude, is interpreted in terms of the spin-orbit torque mechanism. The perpendicular magnetization of the layer is monitored through the anomalous Hall effect measurements. Incorporation of a small amount of heavy Bi atoms into (Ga,Mn)As dilute ferromagnetic semiconductor layer, which causes a substantial increase in the strength of spin-orbit coupling, results in significant lowering the threshold current necessary for magnetization reversal induced by the spin-orbit torque in the tensile-strained (Ga,Mn)(Bi,As) layer. This material may be especially favourable for specific spintronic functionalities utilizing electrically controlled magnetization reversal.

Acknowledgements

K.D. acknowledges financial support from the Foundation for Polish Science through the IRA Program co-financed by the EU within SG OP.